\documentclass[12pt]{article}
\usepackage{epsf,ifthen,amsmath,amssymb,amsfonts}
\usepackage[cp1251]{inputenc}
\usepackage[english]{babel}
\usepackage[colorlinks]{hyperref}
\usepackage{graphicx}
\usepackage{verbatim}
\usepackage{epsfig}
\begin{document}

\title{Robustness of noise-present Bell's inequality violation by entangled state}
\author{Ivan S. Dotsenko
\thanks{E-mail: ivando@ukr.net} \ \ \ \ \ Volodymyr G. Voronov
\thanks{E-mail: ktpist@ukr.net} \\
\it \small Faculty of Physics, Taras Shevchenko Kyiv National University,\\
\it \small 6 Academician Glushkov Ave., 03680 Kyiv, Ukraine}
\date{}
\maketitle

\begin{abstract}
The robustness of Bell's inequality (in CHSH form) violation by
entangled state in the simultaneous presence of colored and white
noise in the system is considered. A twophoton polarization state is
modeled by twoparameter density matrix. Setting parameter values one
can vary the relative fraction of pure entangled Bell's state as
well as white and colored noise fractions. Bell's operator-parameter
dependence analysis is made. Computational results are compared with
experimental data~\cite{Bovino} and with results computed using a
oneparameter density matrix~\cite{Cabello}, which one can get as a
special case of the model considered in this work.
\end{abstract}

\section{Introduction}
After the famous work by Einstein, Podolsky, Rosen (EPR)~\cite{EPR},
where  they expressed the concept of incompleteness of quantum
description of physical reality, there were numerous attempts to
build more complete theories, which would not violate the causality
principle in the classical meaning.

In 1964 Bell~\cite{Bell}, accepting EPR as the  working hypothesis,
formalized it as a deterministic world idea in terms of local hidden
variable model (LHVM) based on the following principles: 1)
measurement results are determined by properties the particles carry
prior to, and independent of the measurement ("realism"); 2) results
obtained at one location are independent of any actions performed at
spacelike separation ("locality"); 3) the setting of local apparatus
are independent of the hidden variables which determine the local
results ("free will")~\cite{Horod}.

Bell showed that the above assumptions impose some  constraints on
statistical correlations in experiments involving bipartite systems.
Such constraints were formulated in the form of the nowadays
well-known Bell's inequalities. Further Bell showed that the
corresponding correlations, which one can obtain by quantum
mechanical rules, violate these inequalities for some quantum
mechanical states called entangled. In this way entanglement is that
feature of the quantum formalism that gives specific purely quantum
correlations that can't be simulated within any classical model.
Later Bell's inequalities were reformulated in the form suitable for
experimental verification or confutation.

In 1982 Aspect's group (Alain Aspect et. al.~\cite{Aspect})
performed  a verification experiment for possible violation of
Bell's inequalities in Clauser-Horne-Shimony-Holt (CHSH)
form~\cite{CHSH}, where a correlation measurement of twophoton
polarization states was provided. Measurement results correspond
well with the quantum mechanics predictions. Experimental data give
Bell's inequality violation by five standard deviations. Numerous
later experiments showed that their results are in agreement with
the quantum mechanical description of nature.

Thus, specific quantum correlations obtained the status of reality
and entangled states, which provide such correlations, became an
object of intensive research. It turned out that entanglement can
play in essence the role of a new resource in such scientific areas
as quantum cryptography, quantum teleportation, quantum
communication and quantum computation. This became a great stimulus
for researching the methods of creation, accumulation, distributing
and broadcasting of this resource.

\section{Noise-present entanglement detection}
One of the most important questions in the considered topic concerns
methods of identifying the presence of entanglement in one or
another realistic quantum mechanical state. So far as entangled
states violate Bell's inequalities, thus, the violation of Bell’s
inequalities can be the basic tool to detect entanglement. In
realistic applications pure entangled states become mixed states due
to different types of noise. Thus a question about robustness of
Bell's inequalities violation against the noise arises. In other
words, one wants to know, under what proportion of an entangled
state and noise in a realistic mixed state can be uncovered the
presence of entanglement. The most reliable source of two-party
entanglement are polarization-entangled photons created by the
parametric down-conversion process (PDC)~\cite{PDC}.

The entangled singlet twophoton state from the PDC process can be
described as a spherically symmetric function, which is one of the
known Bell's states:
\begin{equation}
\label{singlet}
|\Psi^-\rangle=\frac{1}{\sqrt{2}}\left(|01\rangle-|10\rangle\right),
\end{equation}
where $|0\rangle$ and $|1\rangle$ are two mutually orthogonal photon
polarization states. The density matrix for the photon pair state in
the presence of white noise is the following:
\begin{equation}
\label{white_two-photon}
\hat{\rho}_W=p|\Psi^-\rangle\langle\Psi^-|+\frac{1-p}{4}\hat{I},
\end{equation}
where $\hat{I}$ is the $4\times4$ identity matrix. These states are
called Werner states~\cite{Werner}.

Usually in Bell's inequality violation tests with polarization
entangled photons from the PDC Werner states were used. But
experimental evidence and physical arguments show that the colorless
noise model is not good for the description of states received in
the PDC process. A more realistic description is given by an
alternative oneparameter noise model, where a singlet state is mixed
with the decoherence terms, which are called "colored
noise"~\cite{Cabello, Bovino}:
\begin{equation}
\label{colored_two-photon}
\hat{\rho}_C = p|\Psi^-\rangle\langle\Psi^-|+\frac{1-p}{2}
(|01\rangle\langle01|+|10\rangle\langle10|),
\end{equation}

In the polarization density matrices \eqref{white_two-photon}  and
\eqref{colored_two-photon}, varying the parameter $p$ from 0 to 1
one can model different relative proportions for the pure entangled
state $|\Psi^-\rangle$ and the noise. Bell's inequality in the CHSC
form:
\begin{equation}
\label{CHSH}
|\beta|\leq2,
\end{equation}
where
\begin{equation}
\label{Bell_operator}
\ \beta=-\langle A_0B_0\rangle-\langle A_0B_1\rangle-\langle A_1B_0\rangle+
\langle A_1B_1\rangle
\end{equation}
is called the Bell operator.

For maximal Bell's inequality \eqref{CHSH} violation  analysis,
separately in states with white \eqref{white_two-photon} and colored
\eqref{colored_two-photon} noise, in Cabello's work (Adan Cabello at
al.~\cite{Cabello}) the following onequbit observables were taken:
\begin{equation}
\label{A_0}
A_0=\sigma_z
\end{equation}
\begin{equation}
\label{A_1}
A_1=\cos(\theta)\sigma_z+sin(\theta)\sigma_x
\end{equation}
\begin{equation}
\label{B_0}
B_0=\cos(\phi)\sigma_z+sin(\phi)\sigma_x
\end{equation}
\begin{equation}
\label{B_1}
B_1=\cos(\phi-\theta)\sigma_z+sin(\phi-\theta)\sigma_x
\end{equation}

The parameters $\theta$ and $\phi$ in \eqref{A_0}-\eqref{B_1}  are
determining analyzers orientation in experimental devices,
$\sigma_x$ and $\sigma_z$ are the usual Pauli matrices. Computations
showed, that for the Werner state \eqref{white_two-photon} the
maximal value of $\beta$ as $p$-parameter function is the following:
\begin{equation}
\beta_{max}(p)=2\sqrt{2}p
\end{equation}
and for all values of $p$ the maximal value $\beta$ is  obtained by
$\theta=\frac{\pi}{2}$, $\phi=\frac{\pi}{4}$.

Thus, Bell's inequality \eqref{CHSH} is violated only  for
$p>1/\sqrt{2}\approx0.707$. This implies, that in the case, where
the entangled state $|\Psi^-\rangle$ is distorted only by white
noise, entanglement presence can be detected if noise proportion is
less then $\thicksim29\%$.

In the colored noise case \eqref{colored_two-photon} the maximal
value of $\beta$ for different values $p$ is achieved at different
values of angles $\theta$ and $\phi$. The most interesting fact is
that the state \eqref{colored_two-photon} violates the CHSH
inequality for all values $0<p\leq1$. Thus, Bell's inequality
violation is extremely robust against colored noise.

In 2006 the work by Bovino (Fabio A. Bovino et al.~\cite{Bovino})
appeared, discussing the experimental verification of the previously
mentioned predictions concerning CHSH inequality robustness against
colored noise. A crystal (beta-barium borate) was irradiated by a
laser, working in pulsed mode, and in the PDC process photon pairs
in polarization-correlated states were created. These states
correspond to the following polarization density matrix:
\begin{equation}
\label{Bovino_exp}
\hat{\rho} = p|\Phi^+\rangle\langle\Phi^+|+\frac{1-p}{2}
(|00\rangle\langle 00|+|11\rangle\langle 11|),
\end{equation}
where
$|\Phi^+\rangle=\frac{1}{\sqrt{2}}\left(|00\rangle+|11\rangle\right)$
is  one of the four entangled Bell's states. State $|1\rangle$
conforms to ordinary polarization and state $|0\rangle$ conforms to
extraordinary ray polarization in the uniaxial crystal.

Onequbit observables in the Bovino's experiment satisfied the
expressions \eqref{A_0}-\eqref{B_1}. Computations showed, that the
$p$-parameter dependence of the $\beta_{max}$ value in the state
\eqref{Bovino_exp} is just the same as in the state
\eqref{colored_two-photon}. The experimental setup made possible the
regulation of the colored noise fraction, that is parameter $p$ was
varying from zero to almost one. Particular cases in
\eqref{Bovino_exp} are the pure state $|\Phi^+\rangle \ (p=1)$ and
just noise $(p=0)$.

The experimental values of $\beta_{max}(p)$ and the theoretical
predictions comparison showed, that the polarization state
\eqref{Bovino_exp} model generally appropriately describes the
photon pair state from the PDC process. However for all $0<p<1$
experimental values of $\beta_{max}$ were found to be a little
smaller than corresponding theoretical values. Due to that, in
concordance with experimental data, the CHSH inequality is violated
only for $p\gtrsim0.2$, but not for all values of $p$, which
followed from computations. The reason of such discrepancy can be
the presence of some portion of white noise beside colored noise in
the realistic polarization state.

In the current work theoretical analysis for robustness of  Bell's
inequality (in CHSH form) violation with simultaneous presence of
colored and white noise is performed. The density matrix for the
twophoton polarization state in such a generalized model can be
expressed in the form:
\begin{equation}
\label{colored_white}
\hat{\rho}_{CW} = p|\Phi^+\rangle\langle\Phi^+|+\frac{r}{2}
(|00\rangle\langle00|+|11\rangle\langle11|)+\frac{1-(p+r)}{4}\hat{I}.
\end{equation}
Varying the parameter $p$ in the range from $0$ to $1$, one can
change the pure state $|\Phi^+\rangle$ fraction in
\eqref{colored_white}, and putting $r$ from $0$ to $(1-p)$, with the
value of $p$ fixed, one can change relative colored and white noise
fractions. For $r=0$ we have the particular case
\eqref{white_two-photon} (colored noise absence), and for $r=1-p$ we
have \eqref{colored_two-photon} (white noise absence).

In the state \eqref{colored_white} the quantity $\beta$, which
responds to onequbit observables \eqref{A_0}-\eqref{B_1} is a
four-parameter function:
\begin{equation}
\begin{split}
\label{CW_beta}
\beta_{CW}(p,r,\theta,\phi)=cos(\phi)[(2p+r)(sin^2(\theta)+cos(\theta))+{}\\
+rcos(\theta)]-sin(\phi)(2p+r)[cos(\theta)-1]sin(\theta).
\end{split}
\end{equation}
In the colored noise absence $(r=0)$ we have:
\begin{equation}
\beta_W(p,\theta,\phi)=2p\{cos(\phi)[sin^2(\theta)+cos(\theta)]-sin(\phi)[cos(\theta)-1]sin(\theta)\},
\end{equation}
and in the white noise absence $(r=1-p)$:
\begin{equation}
\beta_C(p,\theta,\phi)=cos(\phi)[(1+p)sin^2(\theta)+2cos(\theta)]-sin(\phi)(1+p)[cos(\theta)-1]sin(\theta).
\end{equation}

For fixed values of the  parameters $p$ and $r$ the  expression
\eqref{CW_beta} is a function of $\theta$ and $\phi$. Solving the
extremum problem for two-variable function, one can find the maximal
values $\beta^{max}_{CW}(p,r)$, as well as the angles $\theta$ and
$\phi$, that provide the maximal $\beta_{CW}(p,r)$.

In the Fig.1 the shaded surface graphically displays the
 $\beta^{max}_{CW}(p,r)$ as a function of two variables $p$ and $r$. For comparison
in the figure the plane $\beta=2$ is displayed, which is the
boundary value of Bell's inequality. The surface patch above the
plane $\beta=2$ is the CHSH inequality violation area.

\begin{center}
\noindent \epsfxsize=8cm \epsffile{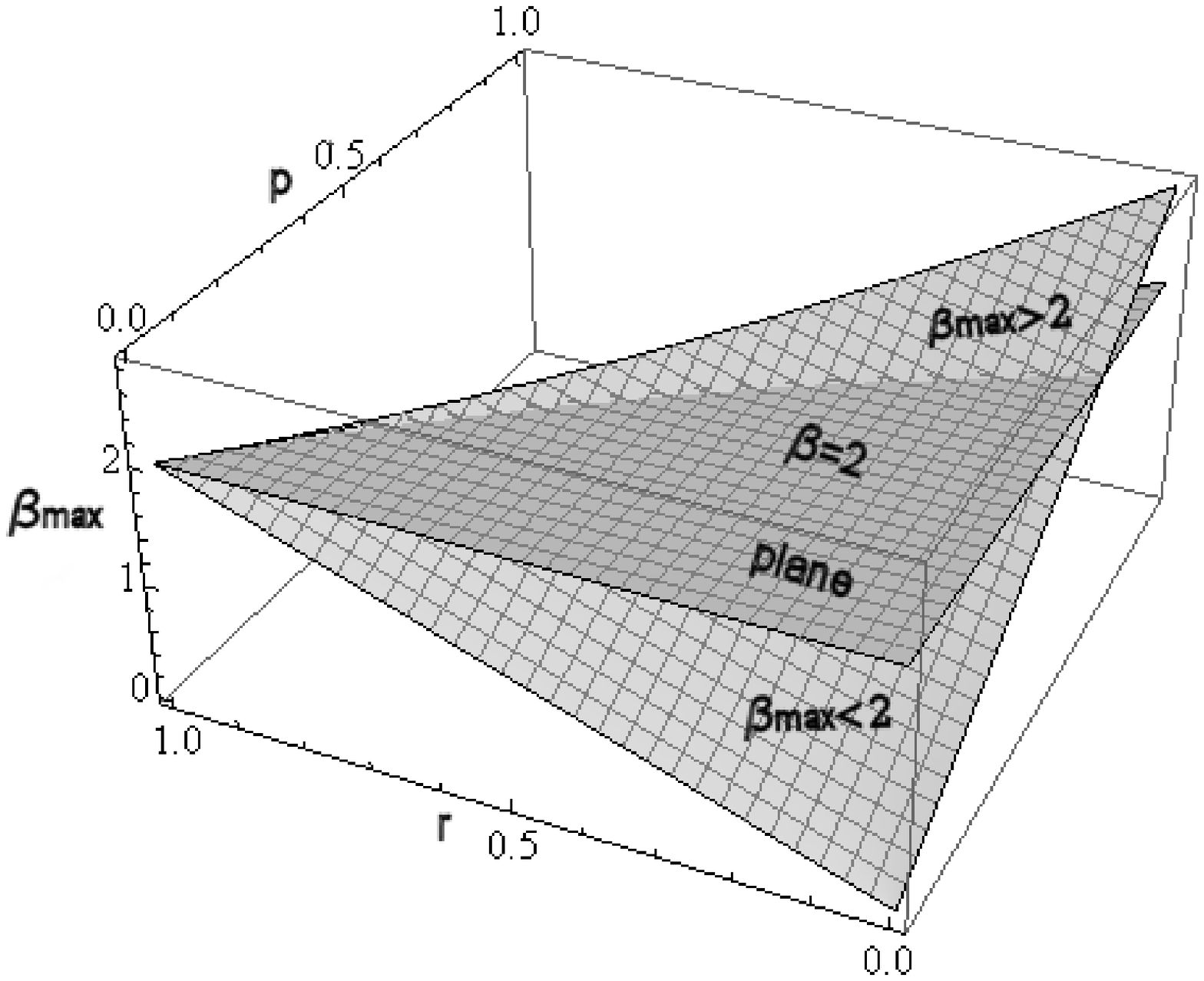}
\end{center}
\noindent{\footnotesize Fig.1. 3D Plot for the maximal Bell operator
values and the $\beta_{CW}=2$ plane, that corresponds to the
classical bound.} \vskip15pt

In the Fig.2 projections on the $(p,r)$ plane of the traces
$\beta=const$ with the surface $\beta^{max}_{CW}(p,r)$ are
represented. From the figure one can see that the straight line
$p+r=1$ (white noise absence) fully lies in the $\beta^{max}>2$
area, which corresponds to the above conclusion, that Bell's
inequality violation is robust against colored noise. For $r=0$
(colored noise absence) Bell's inequality is violated only for
$p>1/\sqrt{2}$. For any fixed $p$ (pure entangled state weight
factor) the $\beta^{max}_{CW}$ decreases with the increasing white
noise fraction. Thus, as expected, adding some amount of white noise
to the colored one can reach better agreement of theoretically
computed $\beta^{max}$ values with experimental ones. Bell's
inequality violation is unsteady under the increasing white noise
fraction for a fixed total amount (white and colored) of noise.

\begin{center}
\noindent \epsfxsize=8cm \epsffile{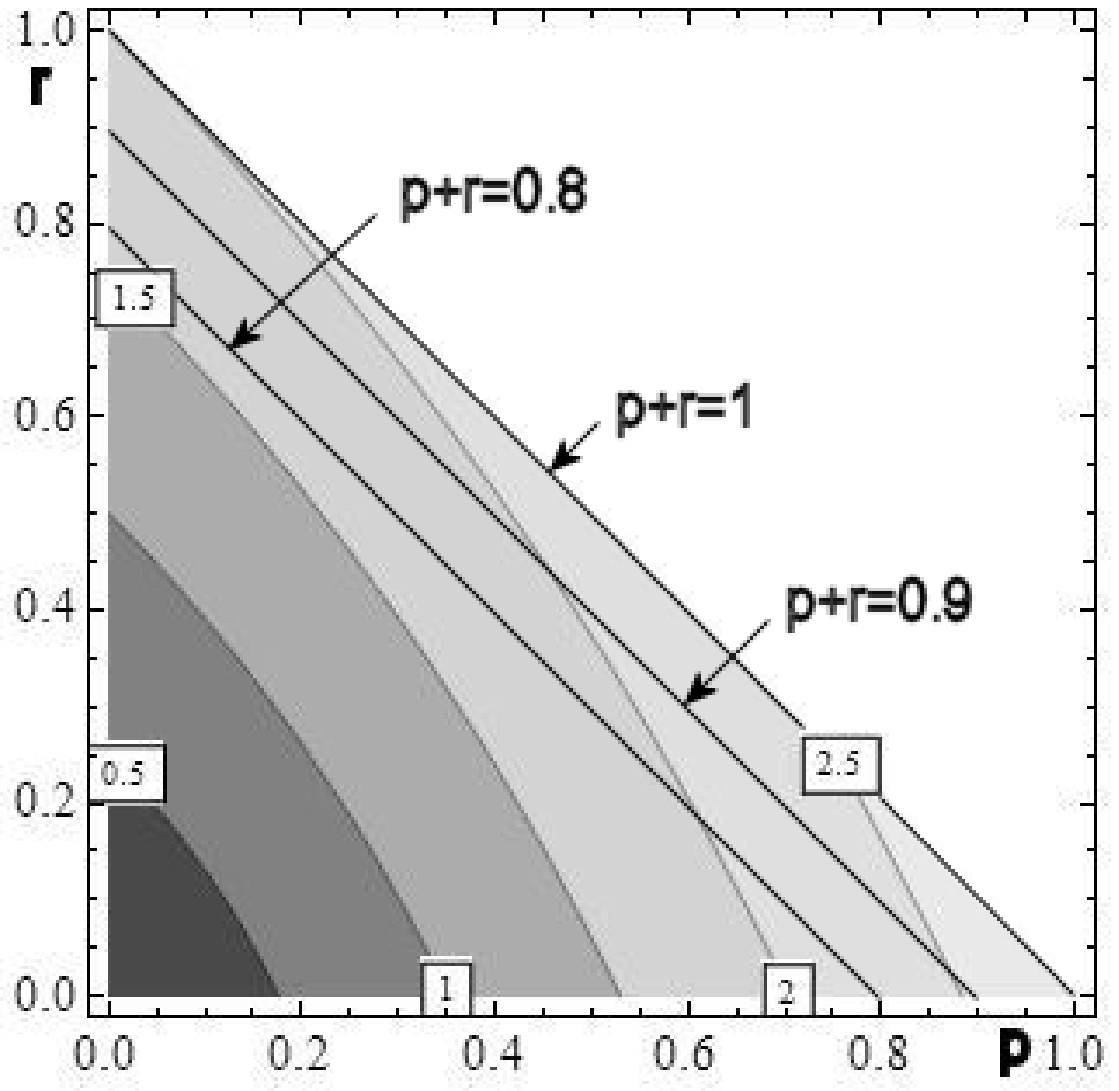}
\end{center}
\noindent{\footnotesize Fig.2. Contour plot for
$\beta^{max}(p,r)=const$ - maximal Bell operator values on
coordinate plane $(p,r)$.} \vskip15pt

\begin{center}
\noindent \epsfxsize=8cm \epsffile{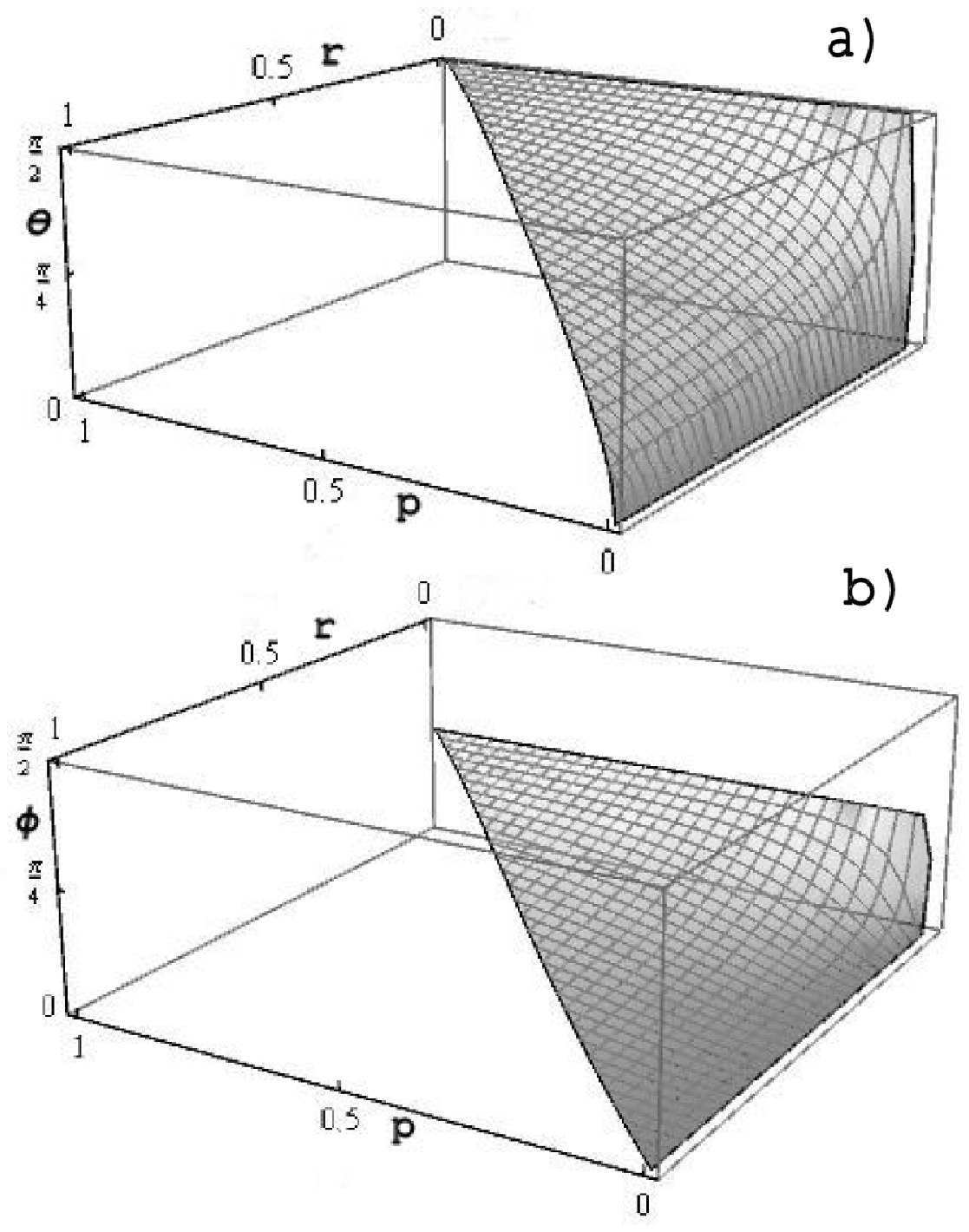}
\end{center}
\vskip-3mm \noindent{\footnotesize  Fig.3.a) Parameter $\theta$
values and b) parameter $\phi$ values, that provide maximal values
of $\beta_{CW}$ for the corresponding parameters $p$ and $r$
values.} \vskip15pt

In the Fig.3a) and 3b) the surfaces, which graphically display the
angles $\theta(p,r)$ and $\phi(p,r)$, that provide maximal values of
$\beta$ as a function of $p$ and $r$ are represented.

In the Fig.4 the curves, which are intersection lines of the
vertical planes $r+p=1$ and $r=0$ with the surface
$\beta^{max}_{CW}(p,r)$ in the case, when in the photon polarization
state there is no white noise, are represented. The dashed line
illustrates the case of colored noise absence. The boundary case
dependencies of $\beta^{max}$ on $p$ and $r$ coincide with the ones
from the work~\cite{Bovino}.

\begin{center}
\noindent \epsfxsize=8cm \epsffile{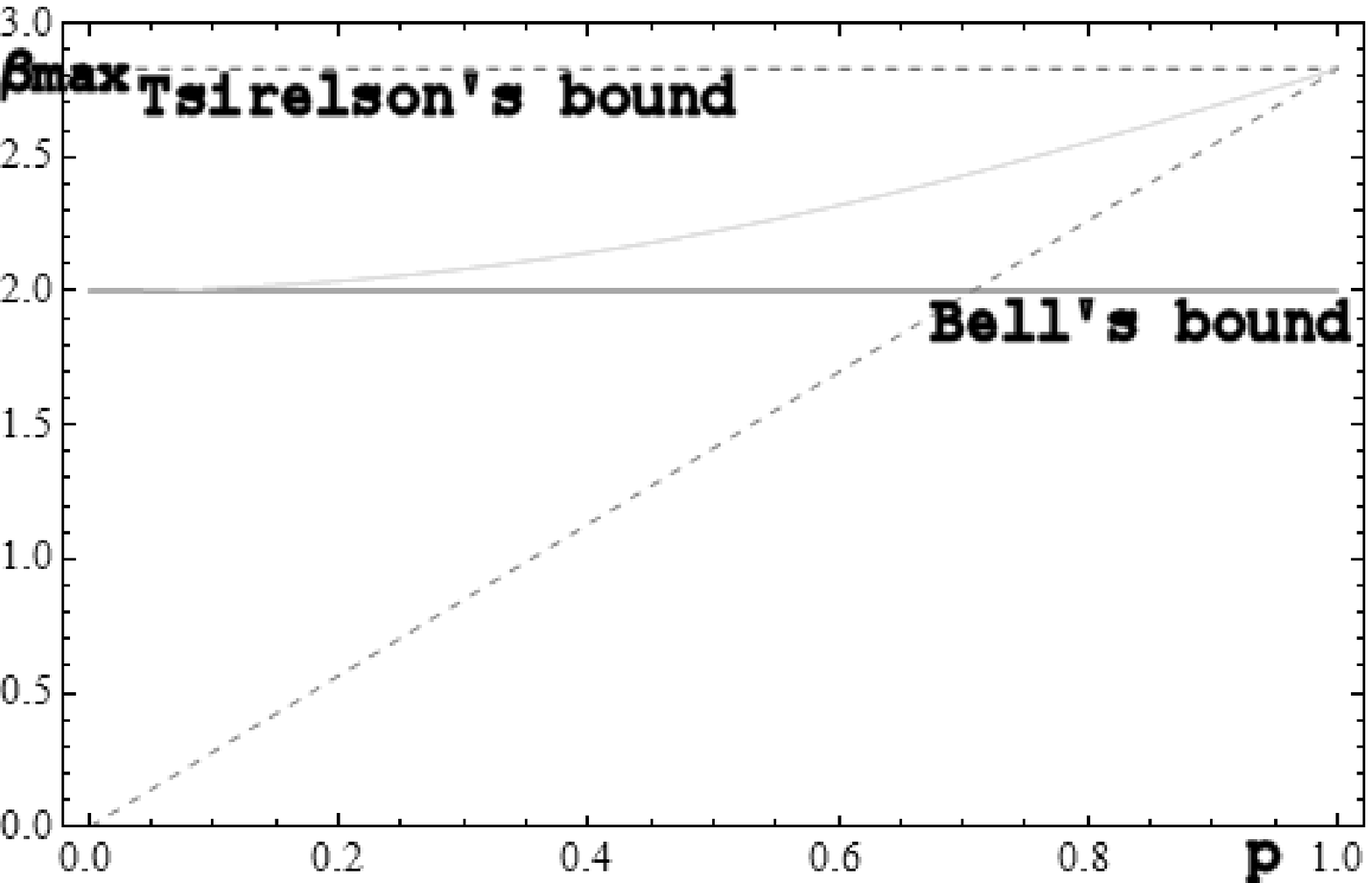}
\end{center}
\vskip-3mm \noindent{\footnotesize  Fig.4. Maximal Bell operator
values in the case $r=1-p$ -- no white noise (top curve) and $r=0$
-- no colored noise (bottom dashed straight line). Classical bound
is $2$. Tsirelson's bound~\cite{Tsirelson} is $2\sqrt{2}=2.83$.}
\vskip15pt

In the Fig.5 the values of the angles $\theta$ and $\phi$, that
provide maximal values of the Bell operator, are represented. Two
solid curves correspond to the case, when in the twophoton
polarization state \eqref{colored_white} white noise is absent
$(p+r=1)$, and two dashed lines correspond to the case, when colored
and white noise enter into the expression \eqref{colored_white} with
the same weight $r=(1-p)/2$. Solid curves coincide with the ones
plotted in the work~\cite{Bovino}. From the figure one can see that
the values of the angles $\theta$ and $\phi$ for a fixed pure
entangled state fraction ($p$ is constant) depend on the
distribution of weighting coefficients of white and colored noise.
Thus, the the orientation of the analyzers for obtaining maximal
values of $\beta$ depends on the fraction distribution between white
and colored noise.

\begin{center}
\noindent \epsfxsize=8cm \epsffile{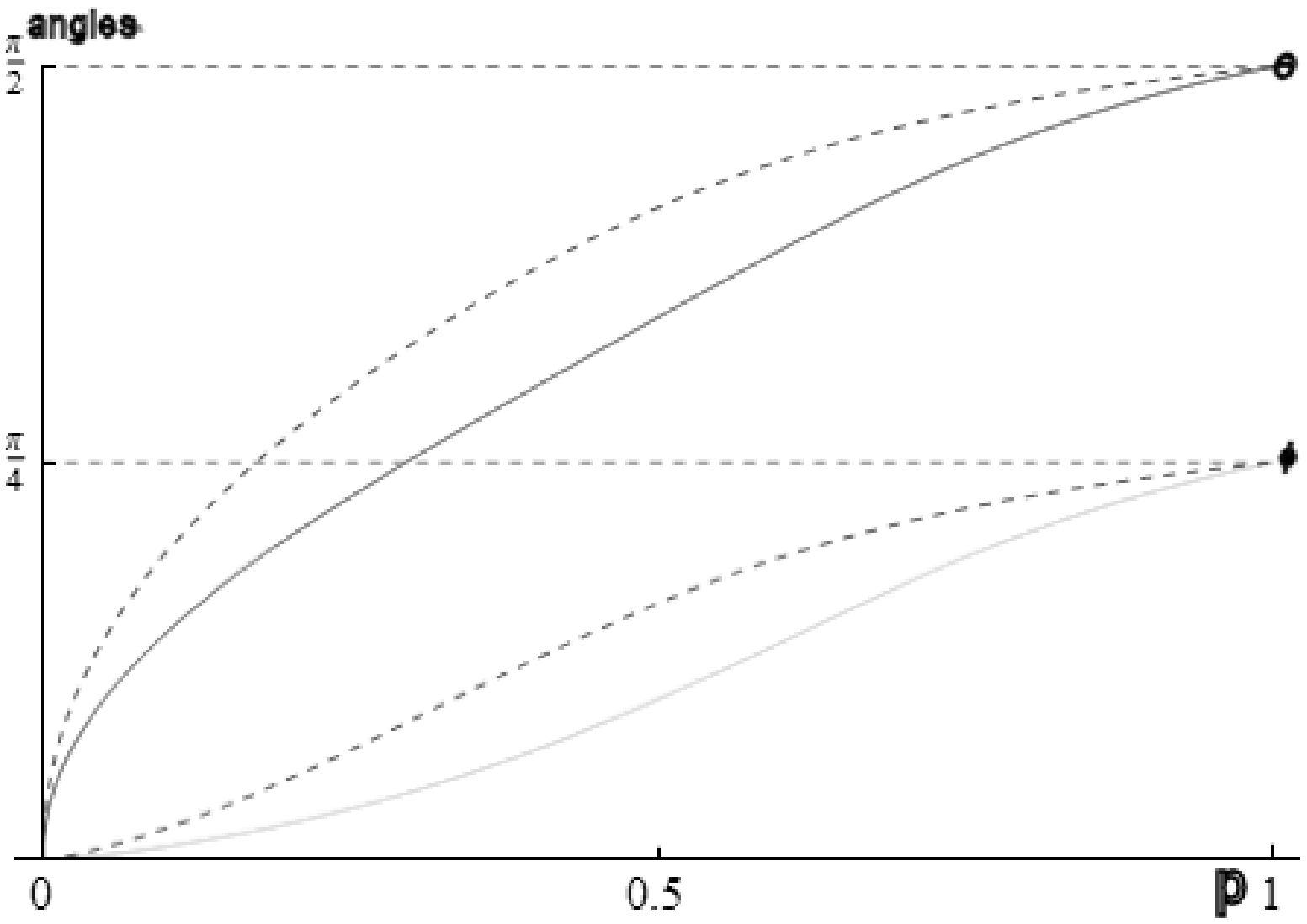}
\end{center}
\vskip-3mm \noindent{\footnotesize  Fig.5. The values of the
parameters $\theta$ and $\phi$, that correspond to the maximal Bell
operator values (two solid curves concern to the case $r=1-p$ -- no
white noise; two dashed curves concern to $r=(1-p)/2$ --- equal
weight coefficients for white and colored noise).} \vskip15pt

In the Fig.6 the points represent the experimental maximal  values
of $\beta$ from the work~\cite{Bovino}; the dashed curve displays
theoretical predictions for the maximal values of $\beta$ on the
oneparameter colored noise model~\cite{Cabello}; the solid curve
illustrates theoretical calculations on the twoparameter
(generalized) noise model with the white noise fraction being
$3.5\%$ of the total noise amount in the system. In the figure we
can see that for such a noise proportion experimental data better
corresponds to theoretical predictions, i.e. the generalized
(twoparameter) noise model is more correct then the oneparameter for
realistic states description. But in this case too, as one can see
in the figure, some experimental points lie above and below the
theoretical curve. According to the twoparameter model, this is
explained by the fact that by moving from one point to other not
only does the total noise amount in the system change, but relative
fractions of white and colored noise does too.

\begin{center}
\noindent \epsfxsize=8cm \epsffile{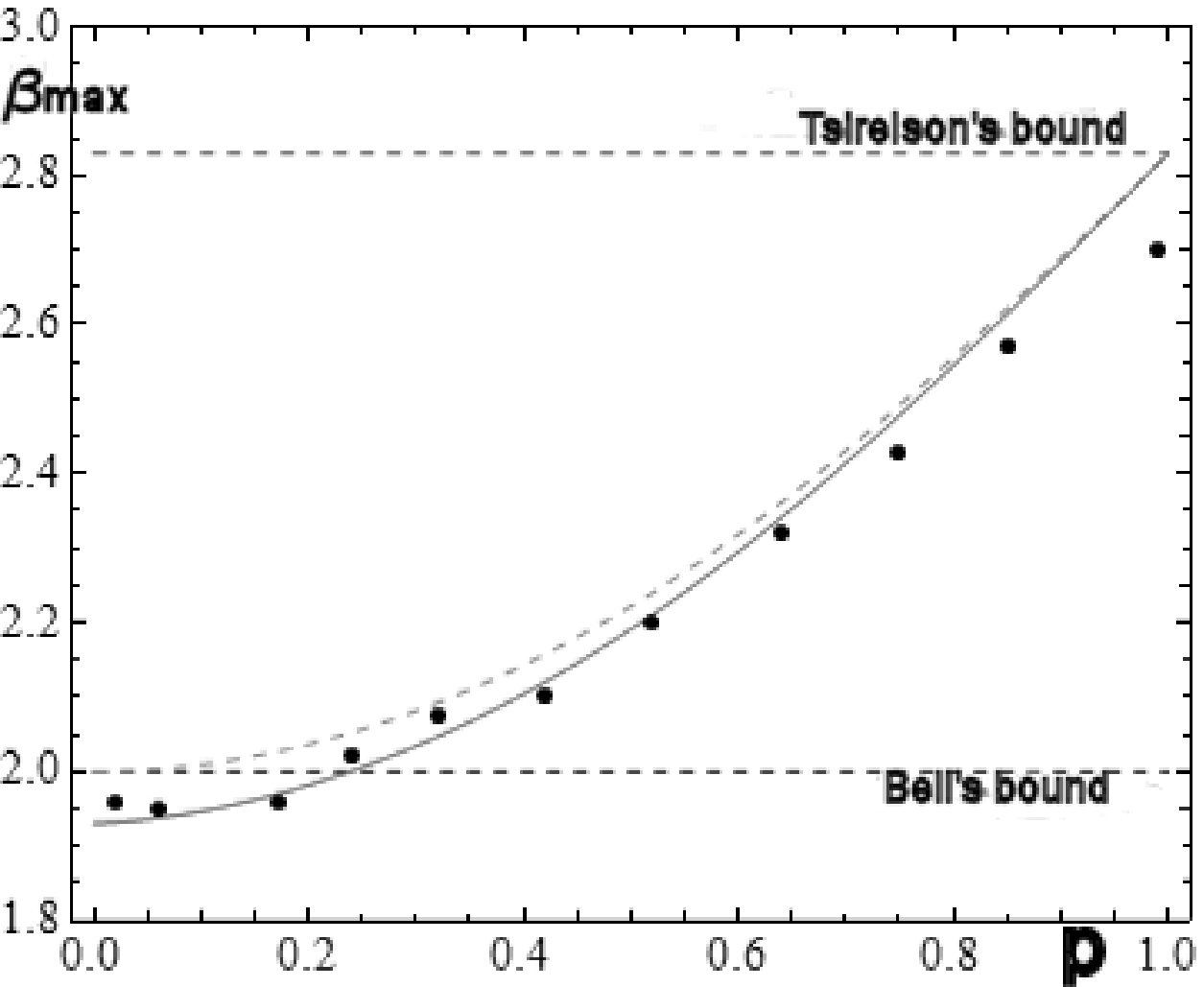}
\end{center}
\vskip-3mm \noindent{\footnotesize  Fig.6. The points represent
experimental maximal values of $\beta$ from the work~\cite{Bovino};
the dashed curve is the theoretical predictions for the maximal
values of $\beta$ in the oneparameter colored noise
model~\cite{Cabello}; the solid curve shows theoretical calculations
in the twoparameter (generalized) noise model with the white noise
fraction being $3.5\%$ of the total noise amount in the system.}
\vskip15pt

This kind of interpretation is absolutely logical, because for the
each measurement experimental setup is tuned up in a new way
(particulary, one has to change the analyzers orientation in space).
Remaining in the theoretical model, which is considered in this
work, and choosing the corresponding parameter $r$ values for each
experimental point (for fixed $p$) one can fully conform theoretical
computations with the experimental data. Let us recall, that the
preselected values of the parameters  $p$ and $r$, according to our
model, determine the pure entangled state fraction and relative
noise fractions. The percentage of white and colored noise
fractions, that give coincidence between theoretical values
$\beta_{max}$ and experimental data, is represented in the table.
Experimental data was taken from the figure in the
work~\cite{Bovino}.

\vskip3mm \noindent{\footnotesize{\bf Table of noise proportions in
the system.  Correspondence with experimental points in the Fig.6}}
\vskip1mm \noindent\
\begin{center}
\begin{tabular}{lccccc}
\hline
 Nr. & $p$ & $1-p$ & white,\% & colored,\% & $r$ \\
\hline
\hline
 1 & 0.02 & 0.98 & 2 & 98 & 0.96\\
 2 & 0.06 & 0.97 & 3 & 97 & 0.92\\
 3 & 0.17 & 0.83 & 4 & 96 & 0.80\\
 4 & 0.24 & 0.76 & 2 & 98 & 0.75\\
 5 & 0.32 & 0.68 & 2 & 98 & 0.67\\
 6 & 0.42 & 0.58 & 5 & 95 & 0.55\\
 7 & 0.52 & 0.48 & 5 & 95 & 0.46\\
 8 & 0.64 & 0.36 & 7 & 93 & 0.40\\
 9 & 0.75 & 0.25 & 15 & 85 & 0.21\\
10 & 0.85 & 0.15 & 15 & 85 & 0.13\\
\hline
\hline
\end{tabular}
\end{center}

\section{Conclusions}
For adequate modeling of the twophoton polarization state, created
in the parametric down-conversion process (PDC type II), one should
take into account the presence of colored noise as well as white.
While Bell's inequality violation is extremely robust against the
colored noise (Bell's inequality is violated for all $0<p\leq1$),
the violation is unsteady under white noise. White noise presence,
that is determined by a weighting coefficient of just $0.1$
$(p+r=0.9)$, as one can see in the Fig.2, leads to Bell's inequality
violation only for $p\gtrsim0.5$. Simultaneously taking into account
both colored and white noise gives possibility to conform
theoretical computations with experimental data. Taking $p$ and $r$
as adjustable parameters one can determine colored and white noise
fractions by comparison of theoretical calculations with
experimental data. The best model is the one which explains
experimental values of $\beta^{max}$ as well as the angles $\theta$
and $\phi$, that provide these values.

\end{document}